# Irreversible Susceptibility of Initial Magnetization Curve

G. Goev[1], V. Masheva[1], J. Geshev[2] and M. Mikhov[1]

[1] Faculty of Physics, "St. Kl. Ohridski" University of Sofia, 1164-Sofia, Bulgaria
[2] Instituto de Física, UFRGS, 91501-970 Porto Alegre, RS, Brazil

*Abstract:* A method for estimation of reversible and irreversible susceptibilities of initial magnetization curves has been developed. It deals only with the energy necessary for magnetizing and demagnetizing the sample, not with the nature of the magnetization processes neither with a specific type of anisotropy, so it could be applied for a wide variety of real materials. A set of minor hysteresis loops of an initially demagnetized sample, plotted with progressively increasing maximum magnetic field, has been used. The obtained results showed excellent coincidence with those calculated by the remanence curve method for a Stoner–Wohlfarth model system.

**Corresponding author:** M. Mikhov, Faculty of Physics, "St. Kl. Ohridski" University of Sofia, 5 J. Bourchier Blvd., 1164-Sofia, Bulgaria; tel. (+359) 2 81 61 877, FAX : (+359) 2 96 25 276; e-mail: mikhov@phys.uni-sofia.bg.

## I. Introduction

One of the basic problems of the physics of applied magnetism is how the magnetic state of a material changes with the field. There are two main types of elementary processes changing the magnetization – reversible and irreversible ones. The irreversible processes are energy consuming and they are responsible for magnetic and thermomagnetic hysteresis as well as for time-dependent magnetization phenomena.

In general, there can be cited two different methods for estimation of irreversible process contribution [1–14] (see also the references in Ref. [11]). The first method uses major hysteresis loop, minor remagnetization curves, and recoil curves [2,7- 8,10]. The other method is based on the IRM and DCD remanence curves, where IRM means the isothermal remanent magnetization and DCD – the remanence measured in direct current demagnetization regime [3-5].

The connection between irreversible susceptibility, $\chi_{\text{irr}}$, magnetic viscosity, $S$, and fluctuation field, $H_{\text{f}}$, i.e. $S = \chi_{\text{irr}} H_{\text{f}}$, is pointed out in a number of papers [1, 3, 4]. $S$ is a function of time $t$, defined through $M(t) = \text{const} + S f(t)$, where $M$ is the magnetization, and $H_{\text{f}} = k_{\text{B}} T / V_{\text{ac}} M_{\text{s}}$ (here $k_{\text{B}}$ is Boltzmann constant, $T$ temperature, and $V_{\text{ac}}$ activation volume [14]). This relation offers a possibility for estimation of irreversible susceptibility from experimentally obtained $S$, provided that $H_{\text{f}}$ is known from independent experiments.

The aim of the present paper is to introduce a variation of the hysteresis loop method for estimation of reversible and irreversible susceptibilities and magnetizations by using initial magnetization curve and symmetrical minor hysteresis loops. The method presented here is not restricted either with the nature of magnetization processes or with a specific anisotropy type. Its validity is proved by comparing the reversible and irreversible magnetizations for a Stoner - Wohlfarth (S-W) model system of disordered single-domain uniaxial particles. It would be noted that recent advances of nanotechnology enable manufacturing of materials consisting of nanosized grains, characterized by magnetization reversal of uniform rotation, very similar to the S-W coherent reversal law [15-18].

**II. The method**

An initial magnetization curve of a virgin sample will be firstly considered, and the corresponding quantities will be indexed by "i". The density of energy $W^i(H)$, associated with magnetizing the sample up to final magnetization $M^i(H)$, is:

$$W^i(H) = \int_0^{M^i(H)} H \, dM^i = \int_0^H H \, \chi^i_{t,\,diff}(H) \, dH, \qquad (1)$$

where $\chi^i_{t,\,diff}(H) = dM^i(H)/dH$ is the total differential magnetic susceptibility. This energy can be expressed as a sum of the energies $W^i_{rev}(H)$ and $W^i_{irr}(H)$, related to the reversible, $M^i_{rev}(H)$, and irreversible, $M^i_{irr}(H)$, magnetizations, respectively:

$$W^i(H) = W^i_{rev}(H) + W^i_{irr}(H) = \int_0^H H\chi^i_{rev}(H) \, dH + \int_0^H H\chi^i_{irr}(H) \, dH. \qquad (2)$$

The irreversible susceptibility can be written as

$$\chi^i_{irr}(H) = \frac{1}{H} \frac{dW^i_{irr}(H)}{dH}. \qquad (3)$$

It can be thus calculated if $W^i_{irr}(H)$ is known. The field dependence of the "irreversible" energy $W^i_{irr}(H)$ may be experimentally obtained from the set of minor hysteresis loops, indexed by "d", by the following procedure, schematically shown in Fig. 1.

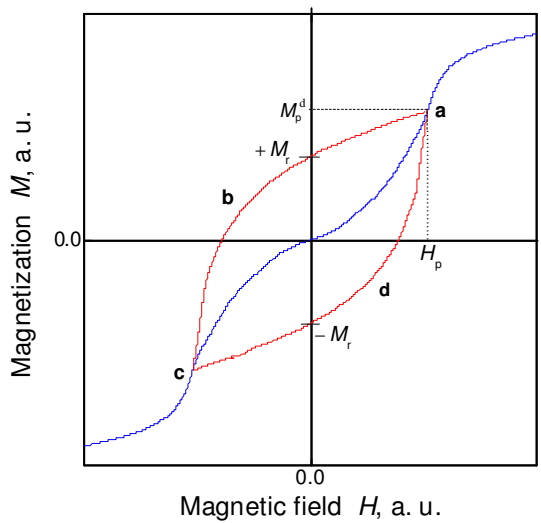

Fig. 1.

Typical initial magnetization curve (blue) and hysteresis loop (red).

The first minor hysteresis loop, $M^d_1(H)$, is plotted with a small enough amplitude $H_1$ of the magnetic field, and the magnetization is changed between $-M^d_1(H_1)$ and $+M^d_1(H_1)$.

Then the amplitude of the field is increased with $\Delta H$ up to $H_2 = H_1 + \Delta H$, and the corresponding magnetization is changed between $-M_2^d(H_2)$ and $+M_2^d(H_2)$. Thus, step by step, the set of minor hysteresis loops, numbered by "p", is plotted until the amplitude of the field reaches a value $H_p$ higher than $H_a$ (where $H_a$ is the anisotropy field), and both branches of the hysteresis loop cannot be distinguished furthermore. $M_p^d(H_p)$'s coincide with the magnetizations of the initial magnetization curve, $M_p^i(H_p)$, measured up to field $H_p$. The magnetic losses of each minor hysteresis loop result from irreversible magnetization processes only. Let consider the arbitrary minor hysteresis loop "p" shown schematically in Fig. 1. The irreversible energy, $W_{irr}^i(H_p)$ which flows from the magnetic field to the sample when its magnetization changes along the initial magnetization curve from point 0 to point a is equal to the irreversible energy connected with the magnetization change along the initial magnetization curve from point 0 to point c. Due to the symmetry of hysteresis loops, the irreversible energy, which flows from the magnetic field to the sample when its magnetization changes between point a and point c is twice the irreversible energy connected with the magnetization from point 0 to point a. This statement, basic for the suggested method, is motivated in more details in the Appendix. Consequently, the irreversible energy for a closed hysteresis loop along a–b–c–d–a trajectory is a sum of four equal amounts of $W_{irr}^i(H_p)$ i.e. the total amount of hysteresis losses, $\Sigma_p$, measured by the area of the considered minor hysteresis loop, is:

$$\Sigma_p = W_{irr}^d(H_p) = 4\,W_{irr}^i(H_p). \tag{4}$$

If the increment $\Delta H$ is small enough ($\Delta H \to 0$), and consequently the number of minor hysteresis loops is large enough ($p \to \infty$), one can write:

$$W_{irr}^i(H) = \frac{W_{irr}^d(H)}{4} \tag{5}$$

Therefore, the irreversible magnetic susceptibility can be calculated from Eqs. (3) and (5). The reversible magnetic susceptibility can be expressed as the difference between the total differential magnetic susceptibility, $\chi_{t,\,diff}^i(H) = dM^i(H)/dH$, and the already obtained irreversible susceptibility, $\chi_{rev}^i(H) = \chi_{t,\,diff}^i(H) - \chi_{irr}^i(H)$.

The reversible and irreversible parts of the magnetization, $M_{rev}^i(H)$ and $M_{irr}^i(H)$, can be obtained by integrating the corresponding reversible and irreversible susceptibilities.

Summing up, the proposed method for estimation of the reversible and irreversible components of the magnetic susceptibility consists of the following steps:
(i) A set of minor hysteresis loops for an initially demagnetized sample is plotted with progressively increasing field amplitude;
(ii) The hysteresis losses $\Sigma_p = W_{irr}^d(H_p)$ are calculated for each minor hysteresis loop, and the irreversible energy $W_{irr}^i(H)$ and susceptibility $\chi_{irr}^i(H)$ are calculated from Eqs. (5) and (3) respectively;
(iii) The initial magnetization curve is built from the edges $(H_p, M_p^d)$ of the minor loops (points a and c in Fig. 1);
(iv) The total differential magnetic susceptibility $\chi_{t,\,diff}^i(H) = dM^i(H)/dH$ is calculated from the initial magnetization curve;

(v) The reversible magnetic susceptibility $\chi^i_{rev}(H) = \chi^i_{t,\,dif}(H) - \chi^i_{irr}(H)$ is then calculated.

## III. Discussion

Unfortunately, we have no a real samples (if any) with well known field dependencies of both reversible and irreversible susceptibilities of initial magnetization curve. Thus, it is impossible to check the validity of the proposed method on real systems, so that we tried to check it on a model one. The method is proved by comparing the reversible and irreversible magnetizations for a system of disordered uniaxial particles, calculated in the framework of the Stoner–Wohlfarth (S–W) model [19], with the ones obtained by the remanence curve method (the IRM curve has been used in the present work) [4].

A normalized initial magnetization curve, $m^i(h)$, and minor hysteresis loops, $m^d_p(h)$, for a S–W model, obtained numerically as described in Ref. [19], are shown in Fig. 2.

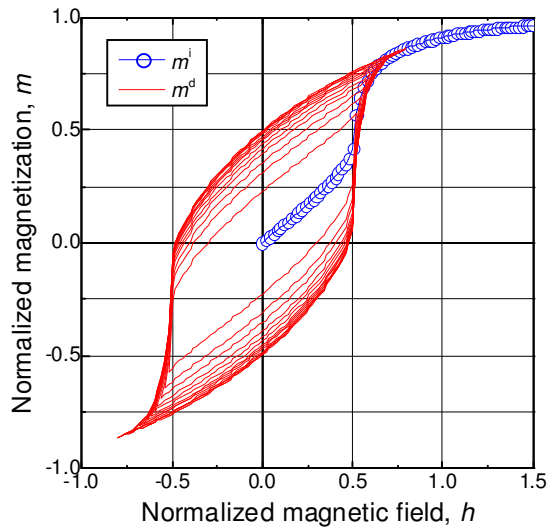

Fig. 2.

Initial magnetization curve, $m^i$ (blue open circles) and set of hysteresis loops, $m^d$ (red solid lines) calculated for an ideal Stoner-Wohlfarth system.

Here $h = H/H_a$ and $m = M/M_s$ are the normalized magnetic field and magnetization, respectively, $H_a$ is the anisotropy field, and $M_s$ is the saturation magnetization. The normalized field has been increased from zero to $h = 1.5$ with an increment of $\Delta h = 0.02$, and in the region of the critical field $h_{cr}$ (i.e., the field at which the first irreversible magnetization changes occur; $h_{cr} = 0.5$ for uniaxial anisotropy), the increment has been reduced to $\Delta h = 0.005$.

Since such a set of hysteresis loops can be easily plotted experimentally in AC regime, the field is assumed to change in the simplest harmonic way:

$$h = h_p \cos(\omega t), \qquad (6)$$

where $h_p$ is the field amplitude for the given minor loop, and $\omega$ is the angular frequency.

It should be noted, that only magnetic losses are considered in the present work, and any time-dependent phenomena as well as eddy current effects are neglected.

In the present work, the parameters of the minor loops and of the initial magnetization curve are obtained by using the Fourier decomposition as described earlier [20-23]. (It should be noted, however, that there are no restrictions concerning the acquisition of these parameters, which can be equally obtained using other methods, e.g., the hysteresis losses can be calculated via the hysteresis loop area technique.) It has been shown [23] that the hysteresis losses $W_{irr}^{d}$ depend only on the amplitude, $\beta_{1,p}^{d}$, of the first sine-harmonic of the Fourier decomposition of the minor hysteresis loop:

$$W_{irr,p}^{d}(h_p) = \pi \beta_{1,p}^{d} h_p m_p. \qquad (7)$$

The field dependence of $W_{irr}^{i}(h)$ is shown in Fig. 3a and the calculated from Eq. (3) field dependencies of the irreversible susceptibility $\chi_{irr}^{i}(h)$, as well as the total differential susceptibility, $\chi_{t,diff}^{i}(h)$, are shown in Fig. 3b.

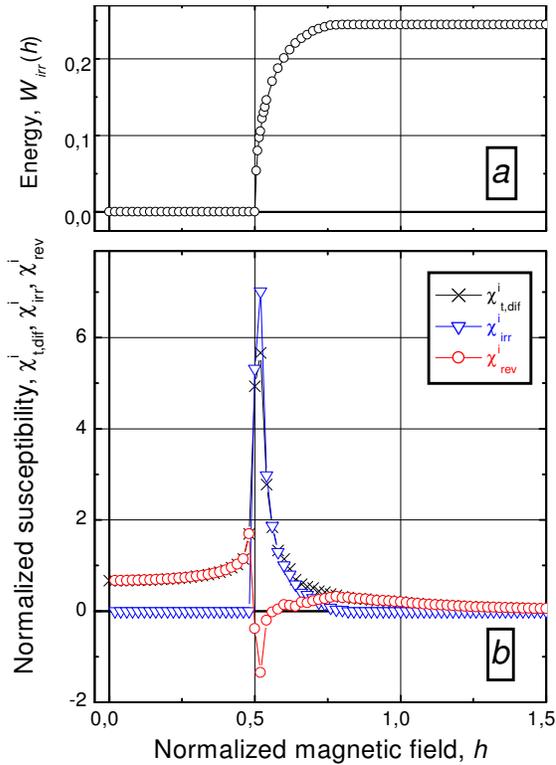

Fig. 3.

(a) Field dependence of irreversible energy;
(b) Field dependencies of: total differential susceptibility, $\chi_{t,diff}^{i}$ (black crosses), irreversible susceptibility, $\chi_{irr}^{i}$ (blue triangles) and reversible susceptibility, $\chi_{rev}^{i}$ (red circles) of initial magnetization curve.

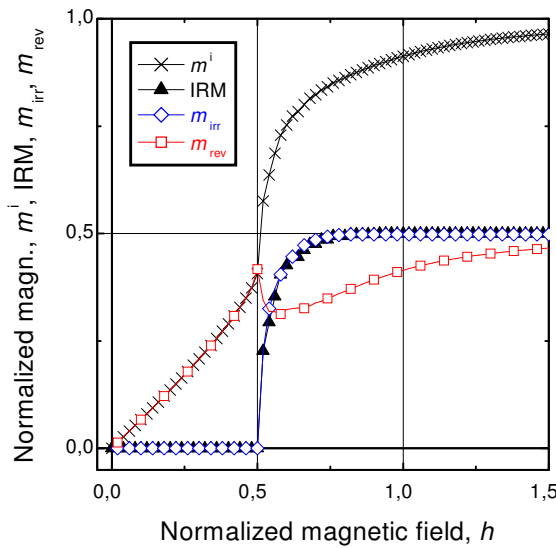

Fig. 4.

Field dependencies of: initial magnetization, $m^{i}$ (black crosses), Irreversible remanent magnetization, IRM (black triangles), irreversible magnetization, $m_{irr}$ (blue diamonds) and reversible magnetization, $m_{rev}$ (red squires).

The irreversible susceptibility $\chi^i_{irr}(h)$ is zero for fields lower than the coercive field, $h < h_c$ ($h_c = 0.479$ for S–W model system with uniaxial anisotropy), and for fields above the anisotropy field, i.e., $h > 1$. In the region of the critical field, the irreversible susceptibility has a maximum that exceeds the value of the total differential susceptibility. The reversible susceptibility has a minimum in the same field region and it even changes its sign.

The calculated here S–W initial magnetization curve $m^i(h)$ and the ones of the irreversible, $m^i_{irr}(h)$, and reversible, $m^i_{rev}(h)$, magnetizations are shown in Fig. 4. It can be seen that the reversible magnetization goes through a maximum at the critical field, which is a direct consequence of the fact that $M^i_{rev}(H)$ depends on $M^i_{irr}(H)$ [11]. For comparison, the IRM is presented in the same figure. There exists a very good coincidence between our results and these obtained by the remanence curves method. A similar field dependence of the reversible magnetization, obtained from DCD curves, is reported in Ref. [11].

The result of the above test is that our method, just like the remanence curves one, works satisfactory for a model S-W system, at least. But since the method is based on energy considerations only it is not restricted either with the nature of magnetization processes or with a specific anisotropy type. In particular, the IRM method in its initial form deals with the coherent magnetization rotation of single domain particles with uniaxial anisotropy. The interpretation of measured remanence plots is strongly dependent on the magnetic anisotropy type [6, 25]. The remanence plots are properly used for the simplest, uniaxial anisotropy case only – for systems with pure cubic or co-existing competing anisotropies there are only a few works that can be found in the literature dealing with remanence curves. When IRM plots are used for estimation of irreversible magnetization the remanence is measured in zero field and the field dependence of irreversible magnetization is more or less suppressed. The use of recoil curves [7, 8] solves this problem. In the present method the field dependence of irreversible susceptibility is omitted. And finally it is easier to measure minor hysteresis loops than remanence curves or recoil curves; in the case of AC measurements it is the only possibility.

### IV. Conclusions

The above described method for estimation of the irreversible susceptibility of initial magnetization curve deals with the energy only, so that it does not consider any specific magnetization mechanisms. It is assumed that the energy used for magnetizing the sample to a state with some remanence is equal to the energy necessary to demagnetize the sample from this state to a state with zero remanence.

The validity of the method is proved on a Stoner–Wohlfarth model system and the calculated irreversible energy, irreversible susceptibility, and irreversible magnetization coincide with the corresponding values, obtained by other methods. It is also shown that the reversible susceptibility changes non-monotonously with the field, even changing its sign.

### Appendix

Initial magnetization curve can be experimentally obtained by the edges of minor hysteresis loops, plotted in AC magnetic field with progressively increasing amplitude, as shown on Fig. 2 [24].

Both reversible and irreversible processes take places during magnetization along initial magnetization curve. They can be characterized by the corresponding magnetizations, $M_{rev}^i$ and $M_{irr}^i$, and magnetic susceptibilities, $\chi_{rev}^i$ and $\chi_{irr}^i$. The upper index "i" means that the parameters refer to the initial magnetization curve. $M_{irr}^i$, $\chi_{irr}^i$, and the energy, $W_{irr}^i$, associated with the irreversible magnetization, are related through:

$$\chi_{irr}^i(h^i) = \frac{dM_{irr}^i}{dh^i} = \frac{1}{h^i}\frac{dW_{irr}^i}{dh^i}. \tag{A-1}$$

The irreversible magnetization, $M_{irr}^i(H^i)$, obtained after increasing the field from 0 to a certain final value, $H^i$, along the initial magnetization curve can be obtained after integrating Eq. A-1:

$$M_{irr}^i(H^i) = \int_0^{H^i} \frac{1}{h^i}\frac{dW_{irr}^i}{dh^i} dh^i. \tag{A-2}$$

Since each minor hysteresis loop is symmetrical, one of hysteresis branches (the bottom one, for example) is enough to be considered.

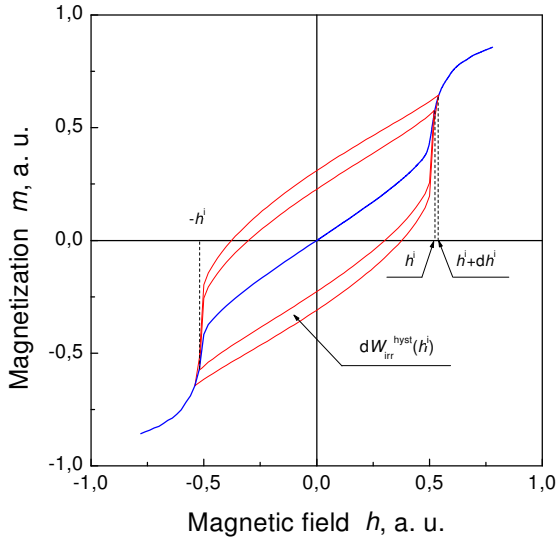

Fig. A-5.

Typical initial magnetization curve and minor histeresis loops, illustrating the consideration given in the Appendix.

The hysteresis losses $W_{irr}^{hyst}(h^i)$, connected to this branch are:

$$W_{irr}^{hyst}(h^i) = \int_{-h^i}^{0} h^d \chi_{irr}^d(h^d) dh^d + \int_{0}^{h^i} h^d \chi_{irr}^d(h^d) dh^d \tag{A-3}$$

The upper index "d" stands for the parameters that are related to the minor hysteresis loop.

After differentiation of Eq. (A-3) one obtains:

$$\frac{dW_{irr}^{hyst}(h^i)}{dh^i} = h^i \chi_{irr}^d(-h^i) + h^i \chi_{irr}^d(h^i) = h^i[\chi_{irr}^d(-h^i) + \chi_{irr}^d(h^i)] = h^i \chi_{irr}^{hyst}(h^i), \tag{A-4}$$

where

$$\chi_{irr}^{hyst}(h^i) = \chi_{irr}^d(-h^i) + \chi_{irr}^d(h^i). \tag{A-5}$$

One should keep in mind that the magnetic susceptibilities $\chi_{irr}^d(-h^i)$ and $\chi_{irr}^d(h^i)$ are different, as seen from the Fig. A-5.

Eq. (A-4) shows how the energy losses increases between the next two bottom branches of minor hysteresis loops, plotted in fields with amplitudes $h^i$ and $h^i + dh^i$ respectively, as shown on Fig. A-6. The increase of energy losses, associated to the above pair of entire minor hysteresis loops is, obviously, twice bigger.

The expression

$$\chi_{irr}^{hyst}(h^i) = \frac{|dM_{irr}^{hyst}(h^i)|}{dh^i} \qquad (A-6)$$

presents half of the increase of irreversible magnetization, associated to the bottom branch of minor hysteresis loop when the field amplitude increases from $h^i$ to $h^i + dh^i$.

By using the Eqs. (A-4) and (A-6) one can obtain the irreversible magnetization in the form:

$$|M_{irr}^{hyst}(H^i)| = \int_0^{H^i} \frac{1}{h^i} \frac{dW_{irr}^{hyst}}{dh^i} dh^i \qquad (A-7)$$

This is the total irreversible magnetization related to the bottom branch of minor hysteresis loop, plotted between $-H^i$ and $H^i$. The necessary condition for both upper and bottom branches of minor hysteresis loops, as well as both branches of initial magnetization curves, plotted in the 1st and 3d quadrants, to be symmetrical for each value of $H^i$ is:

$$M_{irr}^i(H^i) = \frac{1}{2}|M_{irr}^{hyst}(H^i)| .. \qquad (A-8)$$

By making use the Eqs. (A-2), (A-7) and (A-8) one can obtain:

$$\int_0^{H^i} \frac{1}{h^i} \frac{dW_{irr}^i(h^i)}{dh^i} dh^i = \frac{1}{2} \int_0^{H^i} \frac{1}{h^i} \frac{dW_{irr}^{hyst}(h^i)}{dh^i} dh^i$$

$$\frac{dW_{irr}^i(h^i)}{dh^i} = \frac{1}{2} \frac{dW_{irr}^{hyst}(h^i)}{dh^i} \qquad (A-9)$$

Subsequently, $W_{irr}^i(h^i) = \frac{1}{2} W_{irr}^{hyst}(h^i) + const$. Since there is not any hysteresis in low enough magnetic fields, $W_{irr}^i(h^i) = W_{irr}^{hyst}(h^i) = 0$, so that one can accept $const = 0$. Thus, it has been shown that:

$$W_{irr}^i(h^i) = \frac{1}{2} W_{irr}^{hyst}(h^i) .. \qquad (A-10)$$

**Acknowledgments**

This work was supported by Scientific Foundation of the Sofia University, Bulgaria, and partly by the Brazilian agencies CNPq and CAPES.